# Comparison of different deep learning architectures for synthetic CT generation from MR images

Abbas Bahrami[1], Alireza Karimian[2], Hossein Arabi[3]

[1]Faculty of Physics, University of Isfahan, Isfahan, Iran

[2]Department of Biomedical Engineering, Faculty of Engineering, University of Isfahan, Isfahan, Iran

[3]Division of Nuclear Medicine and Molecular Imaging, Department of Medical Imaging, Geneva University Hospital, CH-1211 Geneva 4, Switzerland

**Abstract**

MRI-guided radiation treatment planning is widely applied because of its superior soft-tissue contrast and no ionization radiation compared to CT-based planning. In this regard, synthetic CT (sCT) images should be generated from the patients' MRI scans if radiation treatment planning is sought. Among the different available methods for this purpose, the deep learning algorithms have and do outperform their conventional counterparts. In this study, we investigated the performance of some most popular deep learning architectures including eCNN, U-Net, GAN, V-Net, and Res-Net for the task of sCT generation. As a baseline, an atlas-based method is implemented to which the results of the deep learning-based model are compared. A dataset consisting of 20 co-registered MR-CT pairs of the male pelvis is applied to assess the different sCT production methods' performance. The mean error (ME), mean absolute error (MAE), Pearson correlation coefficient (PCC), structural similarity index (SSIM), and peak signal-to-noise ratio (PSNR) metrics were computed between the estimated sCT and the ground truth (reference) CT images. The visual inspection revealed that the sCTs produced by eCNN, V-Net, and ResNet, unlike the other methods, were less noisy and greatly resemble the ground truth CT image. In the whole pelvis region, the eCNN yielded the lowest MAE (26.03±8.85 HU) and ME (0.82±7.06 HU), and the highest PCC metrics were yielded by the eCNN (0.93±0.05) and ResNet (0.91±0.02) methods. The ResNet model had the highest PSNR of 29.38±1.75 among all models. In terms of the Dice similarity coefficient, the eCNN method revealed superior performance in major tissue identification (air, bone, and soft tissue). All in all, the eCNN and ResNet deep learning methods revealed acceptable performance with clinically tolerable quantification errors.

**Keywords:** Deep learning, MR, synthetic CT, Atlas-based.

## 1. Introduction

Computed tomography (CT) imaging is highly contributive in treatment planning and dose calculation in the radiation therapy (RT) procedure via providing patient-specific 3D electron density maps (attenuation coefficients) [1]. Modern techniques, like the intensity-modulated radiation therapy (IMRT) and volumetric-modulated radiation therapy (VMAT), rely on CT imaging to accurately define and calculate the delivery doses to the target and organs at risk (OAR) [1-3] in their accurate sense. In clinical practice, applying magnetic resonance imaging (MRI) for treatment planning is on an increase due to its zero radiation risk and producing high-contrast soft-tissue images compared to CT. Some clinical studies revealed that the functional MRI information, like the diffusion-weighted imaging (DWI) and dynamic contrast-enhanced imaging, is highly contributive in identifying the active tumor sub-volumes in head and neck cancers [4, 5].

Currently, the complementary information in MR and CT images is exploited through the deformable image alignment for precise delineation of the target volumes and generating the electron density maps [2, 4]. In this context, the errors associated with the MR/CT image alignment introduce a systematic uncertainty in organ delineation and dose calculation which is more outstanding in small tumors or complex organs at risk (OAR) [6, 7]. To fully benefit from the merits of MR imaging in radiation therapy workflow, the MRI-only RT, which solely relies on MR images for organ delineation and electron density map creation is introduced. The MRI-only RT eliminates the need for CT imaging, which decreases the number of scans and the associated costs, next to, reducing the received dose particularly for the patients requiring multiple scans during the treatment process [2, 6, 8]. The MRI-only RT faces certain challenges like the geometric distortions due to magnetic field non-uniformities, absence of cortical bone signal in the MR images, and estimation of the accurate electron density map. The primary challenge in MRI-only RT is that the MR signals correlate to the tissue proton density and relaxation properties rather than tissue attenuation coefficients unlike the CT images [9-11]. The same challenge is evident in the simultaneous PET/MR (and SPECT/MR) systems for the task of PET attenuation correction [12-16]. Approaches in generating synthetic (pseudo) CT from MRI data have been and are being proposed [8, 17, 18].

The methods adopted in generating synthetic CT images consist of tissue segmentation-based [19], Atlas-based [20, 21], and AI-based [14, 22, 23] categories. The first approach generates attenuation maps by segmenting the MR images into a couple of major tissue classes followed by assigning predefined attenuation coefficients to each tissue class [16]. Discrimination of the bony structures from the air is challenging in conventional MR imaging due to their very weak and similar signals [10]. Applying ultra-short echo time (UTE) and zero-echo-time (ZTE) could address this issue, while, these MR sequences suffer from long acquisition time and low signal to noise ratio (SNR) [24].

The second approach consists of deformable image registration algorithms for aligning the target MR image to the corresponding MR atlas images. The one-to-one correspondence of the MR and CT atlas images allows the CT images in the atlas dataset to be applied in estimating the synthetic CT for the target MR image [7, 25]. Their performance highly depends on the availability of similar anatomical and/or pathological variations in the atlas dataset, thus, their major drawback [3, 15].

Recently, deep learning techniques have exhibited high performance in image segmentation, denoising, reconstruction, and, image synthesis in specific [26-31]. Convolutional neural networks (CNN)s, as a multi-layer model of interconnected neurons, can learn the complex non-linear mapping from MR to CT images to synthesize patient-specific pseudo-CTs. In this context, Han [32] developed a CNN (a U-net model with conventional encoder/decoder parts for generating synthetic CTs (sCT) MR images of the brain. This model, trained by the 2D MR and the corresponding CT slices, performed well over the Atlas-based method in terms of accurate CT value estimation. Nie et al. [33] suggested a 3D fully convolutional neural network to generate pseudo-CTs from MR images by applying a patch-wise training scheme of a context-aware generative adversarial network (GAN). Quantitative analysis of the generated sCTs revealed the superiority of the proposed method over other deep learning models and the Atlas-based method. Wolterink et al. [34] developed a novel CycleGAN model to synthesize CT images of brain by applying a dataset of unpaired MR and CT images with promising outcomes in comparison with the conventional CNN model.

There exist many promising deep learning architectures proposed for the task of sCT generation from MR images. Though these models have been and are being properly evaluated, there exist no comprehensive studies to compare/evaluate the different deep learning models using the same dataset. Such a study would shade more light on the efficiency, strength, and shortcomings of these models as to their appropriate adoption in clinical or research settings.

The objective of this study is to compare some popular, state-of-the-art deep learning architectures (including eCNN, U-Net, GAN module, V-Net , ResNet) where the same dataset and evaluation metrics are applied. An Atlas-based synthetic CT generation approach is implemented to provide the baseline for assessing the performance of the deep learning models.

## 2. Materials and Methods

### 2.1 Data acquisition and pre-processing

A dataset of 20 co-registered MRI and CT images of the male pelvis, who referred to the department of radiation therapy for the treatment of prostate cancer were employed in this study. These patients' age range is within 56 and 67 years, (68±3) with a BMI of 18.9 to 34.8 kg/m$^2$ (25±2.5). CT images were acquired on a GE LightSpeed RT (Milwaukee, USA) in a $256 \times 256 \times 128$ matrix with a $1.5625 \times 1.5625 \times 2$ mm$^3$ voxel size. All CT scans are taken on the patients' empty rectum and a full bladder. The MR images were generated through the Siemens Skyra 3T scanner (Erlangen, Germany) where a 3D T2-weighted 1.4 mm isotropic sampling perfection with application-optimized contrast covering the whole pelvis area is applied. The MRI voxel size is originally $1.4 \times 1.4 \times 2$ mm$^3$ that is converted into the corresponding CT image resolution after co-registration. A combination of rigid and non-rigid transformations based on the normalized mutual information loss function and B-spline transformation (implemented in the Elastix[1] package) was applied for MRI to CT image registration. The maximum time interval between CT and MR imaging was one day. Before image registration, the intra-subject non-uniformity of MR intensities was corrected by applying the N4 ITK software followed by denoising through a bilateral edge-preserving filter [35, 36]. The inter-patient MRI intensity variation was reduced by matching the histogram to a common histogram template.

**2.2 Synthetic CT generation approaches**

**2.2.1 ResNet architecture**

Deep residual networks formed through stacking several residual blocks, were introduced by He *et al.*, where the gradient vanishing problem within the training of the deep neural networks and reducing computational cost are addressed [37]. Residual or shortcut connections with an identity map lead to skipping one or more layers in a network (Figure 1).

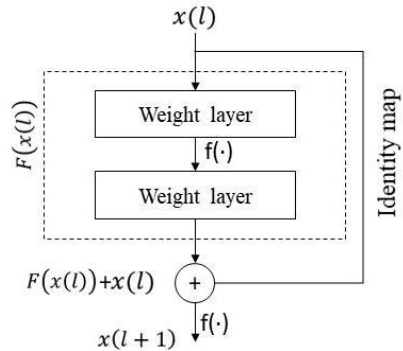

**Figure 1.** Residual block including residual function (F) and identity map [37].

The residual function (F) yields an output as:

$$F(x(l)) = w_{i2}.[f(w_{i1}.x(l) + b_{i1})] + b_{i2} \tag{1}$$

where, $x(l)$ is the input data, $f$ is the activation function, and $w_{i1}$ and $w_{i2}$ are the trainable parameters, and $b_{i1}$ and $b_{i2}$ are the corresponding biases. The identity mapping introduces no extra parameter and computational complexity to the model. The residual block input in Figure 1 is added directly to the output, thus, the output of a residual block in the l-st layer expressed as:

$$x(l+1) = f(F(x(l)) + x(l)) \tag{2}$$

The residual connections enable the direct propagation of signals in the forward and backward paths from one block to the others. Implementing residual connections within the training of a network reduces the border effects of the convolution process which leads to a decrease in distortion near the borders.

---

[1] http://elastix.isi.uu.nl/

The architecture of the ResNet model (Figure 2), consists of 20 convolutional layers where every set of two convolutional layers is stacked by the residual connections. Each convolutional layer is composed of an element-wise rectified linear unit (ReLU) and a batch normalization (BN) layer. In the initial layers, 3×3×3 filters designed to detect/extract low-level image features are applied. For extracting mid-level and high-level image features, the kernels are dilated with factors of two and four in the deeper layers. The output of the final layer, a fully connected softmax layer, has dimensions equal to that of the input image [38].

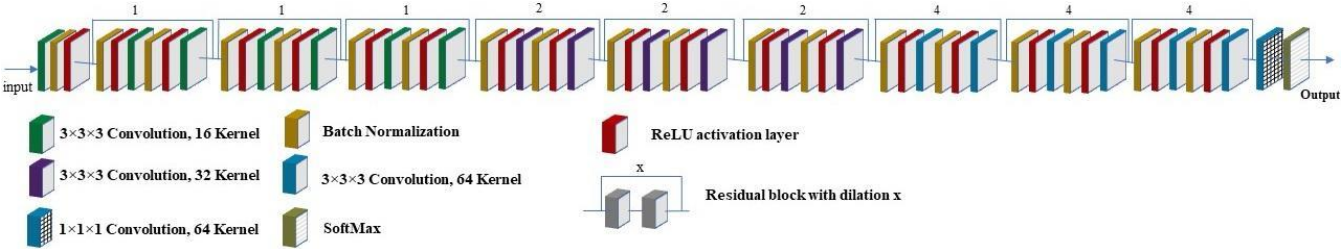

**Figure 2.** Architecture of ResNet model [38].

**2.2.2 eCNN achitecture**

The efficient CNN (eCNN) model was developed based on the conventional encoder-decoder networks in the U-Net model, where, some modifications were made to extract discriminative image features from the input MRI for generating accurate sCTs [39]. In the eCNN model, each simple plain U-Net model convolutional layer is replaced with the building structures proposed by He *et al.* [37]. According to the residual block in Figure 1, the building structure is designed by applying two 3×3 convolutional layers, where each layer is followed by batch normalization and SeLU activation layers to avoid the dying rectified linear unit (ReLU) or dead state. An identity shortcut connection inside the building structure transfers some information from upper to lower layers (serving as a residual connection). Another set of batch normalization and SeLU activation layers is inserted to complete the building structure as illustrated in Figure 3.

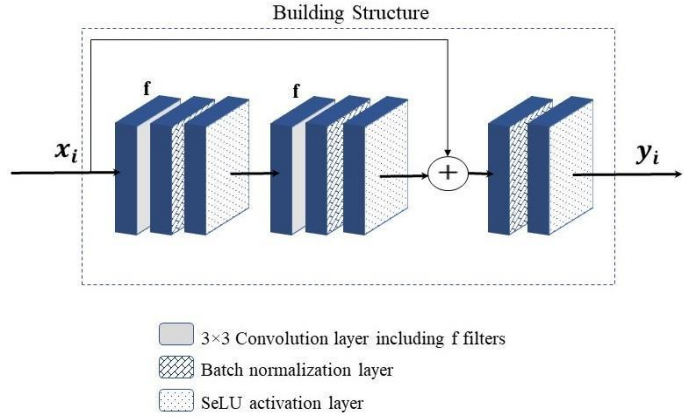

**Figure 3.** The building structure in the eCNN model. *f* denotes the number of filters in each convolutional layer.

The number of filters in each unit of the building structures is the same as those of the corresponding convolutional layers in the U-Net model. The upsampling layers were also replaced with the trainable deconvolutional layers. Moreover, the max-pooling indices and high-resolution feature extraction modules were added to the eCNN model for efficient training performance. The overall architecture of the eCNN model [39] is shown in Figure. (4).

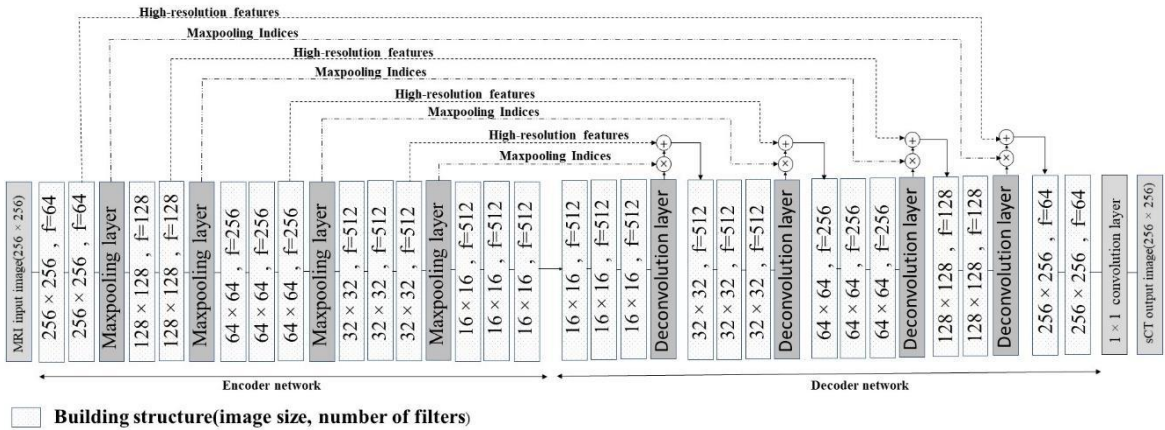

**Figure 4.** The architecture of the eCNN model. The dashed arrows labeled with high-resolution features indicate the connections between encoding and decoding blocks to transfer high-resolution features. The dashed arrows with max-pooling are the connections between the max-pooling layers and decoding networks for transferring the corresponding indices

### 2.2.3 GAN architecture

The general adversarial networks (GAN)s were suggested by Goodfellow *et al.* [40]. This network consists of two adversarial generative and discriminative models that are trained simultaneously [22, 29]. The generative model learns to generate new data, while the discriminator model determines the probability of whether the input data is real or synthesized by the generator. Samples are generated through passing random noise across a multilayer perceptron generator and the output will be fed into the disciminator. The discriminator output would be a scalar likelihood classifying the input as true or false. The discriminator is trained to enhance the differentiation ability, the generator is trained to maximize the probability of the discriminator by assigning an incorrect label to the synthetic data.

The GAN model proposed by Hu et al. [41] is implemented in this study, where, the generator component exploits a random Gaussian noise distribution with zero mean and unit variance and a ReLU activation function. To generate the synthetic images of proper size, the up-scaling layers which are composed of a transposed convolution with 2×2 stride, and convolution with BN and ReLU are applied. These layers duplicate the previous feature maps' size and halve the number of channels. The final layer is of two parts: a convolution kernel, with BN and ReLU, and a convolutional operator, with hyperbolic tangent function without BN to maintain the true statistical features of the data [30].

The discriminator network takes both synthetic and real images as the inputs, where, the first convolutional layer with a 5×5 kernel size and leaky ReLU (LReLU) as activation function, forms the initial feature maps of the same size as the input image. The ResNet blocks and down-scaling layers duplicate the number of channels and halve the size of feature maps in each layer. Each ResNet layer has two convolution kernels, both with BN and LReLU, and each down-scaling layer has a 2×2 stride convolution with BN and LReLU. The Logit component is inserted after the final layer that consists of a ResNet, a projection before and after ReLU. Except for the first layer in the discriminator, all the convolutional layers in the generator and discriminator apply kernels with a size of 3×3 [41]. The overall structure of the GAN model is shown in Figure 5.

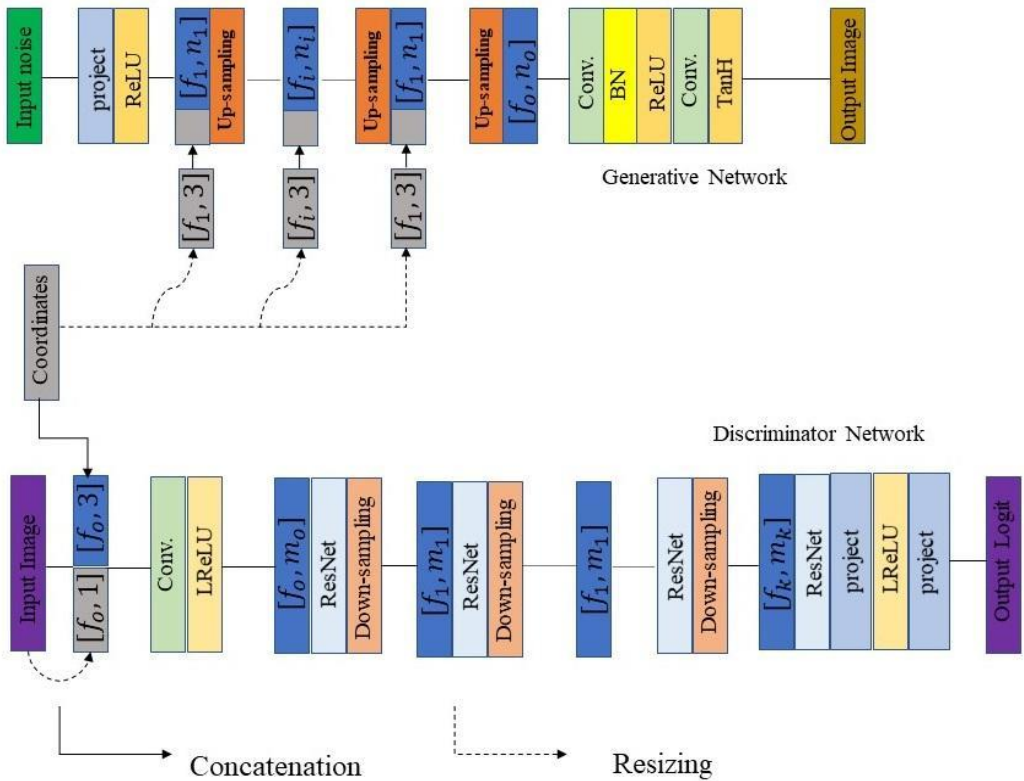

**Figure 5.** The architecture of the GAN model proposed by [41].

## 2.2.5 V-net architecture

Next to the ResNet and GAN models, the V-net structure [42], a 3D fully convolutional neural network, is implemented/evaluated in this study. The core structure of this model consists of a compression path and a decompression, wherein both sides mirror each other. The compression side is divided into different blocks/stages which treat the input data at different resolutions. Each/every stage/block consists of 1-3 convolutional layers. This stage/block, first, learns a residual function where, the input of each stage is processed through a non-linear function and next, is added/linked to the last convolutional layer of the same stage. This model was originally proposed for the semantic image segmentation task [42], and due to its promising performance, it was applied in image regression applications. In this study, the V-net is implemented with an L2-norm loss function to synthesize CT images from MR images.

## 2.2.6. U-net architecture

The early U-net architecture was introduced by Ronneberger et al. in 2015 as one of the fundamental architectures of the deep learning approaches with robust and promising performance in many applications [43]. Fully convolutional layers in the contracting and expanding parts are applied in this model, (Figure 6). The basic U-net model was implemented in this study to be compared with other advanced deep learning models. The L2-loss function led to the peak performance of this model in synthesizing the CT images from MR images.

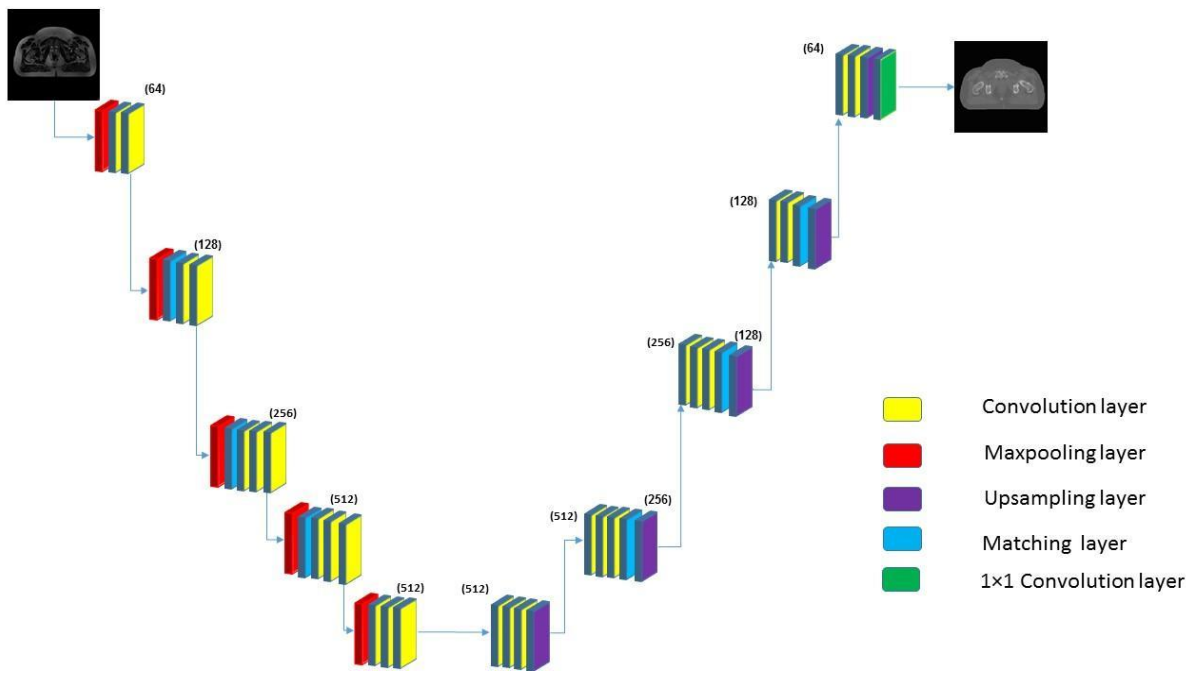

**Figure 6.** The architecture of the U-net model.

## 2.2.4 Atlas-based method

The Atlas-based method is considered a robust and effective approach for synthetic CT image generation from MRI [3]. An Atlas-based sCT generation method is implemented in this study to provide a baseline for the deep learning approaches' performance assessment. The Atlas-based approach includes pairwise registration of the entire MR images (19 out of 20 MRI images) in the dataset into the target MR image following a leave-one-out cross-validation scheme. A combination of the rigid and non-rigid deformation based on the B-spline transform function and normalized mutual information loss function (implemented in Elastix package) were applied in image registrations to align the MR images. Given the transformation maps obtained from the MR image alignment, the corresponding CT images were transformed to the target MR image coordinate. The voxel-wise average of the aligned altas CT images to the target MR images was calculated to create an sCT image for the target MR image.

## 2.3   Evaluation strategy

In evaluating the synthetic CT images generated by the different approaches, the patients' CT images constitute the reference. The entire CT images (reference and synthetic CT images) were segmented into the major tissue classes of air, bone, and soft tissue. The intensity threshold of <-400 HU, and >160 HU were assumed for air and bone segmentation, respectively. The voxels within the -400 to 160 HU range were considered as the soft tissue. To assess the identification accuracy of the major tissue classes in the generated sCT images, the Dice similarity coefficient (DSC) was calculated for each segmented region as follows:

$$DSC(A_r, A_S) = \frac{2\,|A_r \cap A_s|}{|A_r| + |A_s|} \tag{3}$$

where, $A_r$ and $A_S$ represent the volume of a specific tissue in the reference CT images and synthetic CT images, respectively. The mean absolute error (MAE), mean error (ME), Pearson correlation coefficient (PCC), structural similarity index metric (SSIM), and peak signal-to-noise ratio (PSNR) metrics were computed within different major tissue classes for different sCTs as follows:

$$ME = \frac{1}{N}\sum_{i=1}^{N} dA(i) \tag{4}$$

$$MAE = \frac{1}{N}\sum_{i=1}^{N} |\,dA(i)\,| \tag{5}$$

$$PCC(CT, sCT) = \frac{\sum_{i=1}^{N}(CT(i)-\overline{CT})(sCT(i)-\overline{sCT})}{\sqrt{\sum_{i=1}^{N}(CT(i)-\overline{CT})^2}\sqrt{\sum_{i=1}^{N}(sCT(i)-\overline{sCT})^2}} \tag{6}$$

$$PSNR = 10\log\left(\frac{I^2}{MSE}\right) \tag{7}$$

$$SSIM = \frac{(2\mu_r\mu_s + K_1)(2\delta_{rs} + K_2)}{(\mu_r^2 + \mu_s^2 + K_1)(\delta_r^2 + \delta_s^2 + K_2)} \tag{8}$$

where both $A_s(i)$ and $A_r(i)$ are the intensity of i[th] voxel in the sCT and reference CT images and $dA(i) = (A_s(i) - A_r(i))$, N is the number of voxels for each of the tissue classes. In Eq.6, the $\overline{CT}$ and $\overline{sCT}$ are the means of reference CT and synthetic CT images, respectively. In Eq. (7), $I$ is the maximum intensity value of the reference CT or synthetic CT images, and MSE denotes the mean square error. In Eq. (8), both $\mu_r$ and $\mu_s$ are the mean intensity value, and $\delta_r$ and $\delta_s$ are the variances of the CT images. Parameters $K_1 = (k_1 I)^2$ and $K_2 = (k_2 I)^2$ with the constants of $k_1 = 0.01$ and $k_2 = 0.02$ were defined to avoid the division with small denominators.

The above-mentioned metrics were measured over the entire pelvis area to assess the overall performance of the different synthetic CT generation approaches.

## 3. Results

The cross-sectional views of synthetic CT images generated by the different approaches (eCNN, U-Net, GAN, V-Net, ResNet, and Atlas-based method) next to the corresponding MR and reference are shown in Figure (7).

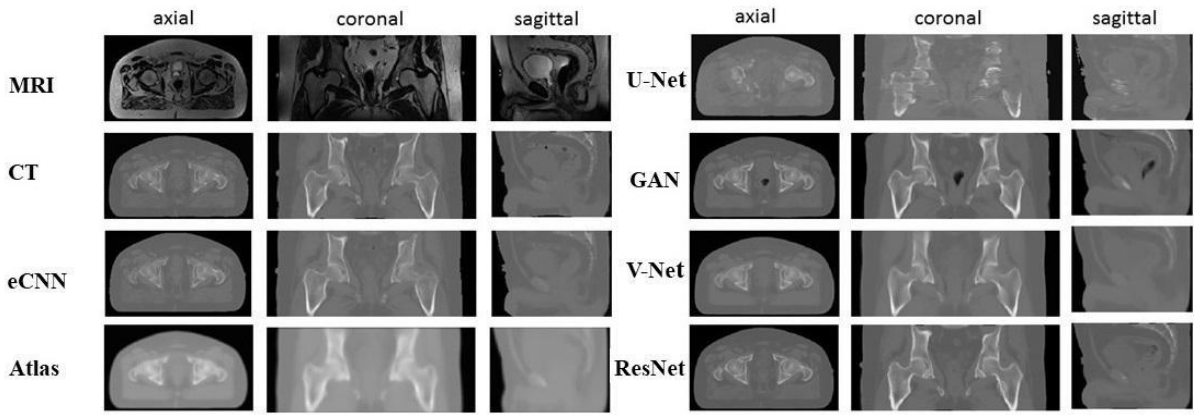

**Figure 7.** Qualitative comparison of sCT images generated by the different approaches together with the reference CT and the target MR images.

It is observed that the sCT images generated through the eCNN, V-Net, and ResNet models are less noisy and correspond well with the reference CT.

The mean and standard deviation of MAE, ME, PCC, SSIM, and PSNR metrics computed on the whole pelvis region to evaluate the different methods' performance in comparison to the ground truth CT images over the 20 test subjects are tabulated in Table 1.

**Table 1.** Statistics of image comparison between the reference CTs and synthetic CTs generated through the different methods on whole pelvis region in terms of MAE, ME, PCC, SSIM, and PSNR (average ± standard deviation).

|         | eCNN        | Atlas-based  | U-net        | GAN         | V-net        | ResNet       |
|---------|-------------|--------------|--------------|-------------|--------------|--------------|
| **MAE(HU)** | 26.03±8.85  | 66.31±20.58  | 43.73±8.33   | 37.26±5.76  | 53.93±13.06  | 31.07±7.42   |
| **ME(HU)**  | 0.82±7.06   | 6.29±38.59   | 8.27±11.45   | 0.55±12.57  | -11.69±20.02 | -3.83±11.48  |
| **PCC**     | 0.93±.05    | 0.83±.05     | 0.78±.06     | 0.89±.02    | 0.86±0.02    | 0.91±0.02    |
| **SSIM**    | 0.98±.01    | 0.95±.02     | 0.95±0.01    | 0.97±0.01   | 0.97±0.01    | 0.98±0.01    |
| **PSNR**    | 27.12±3.51  | 23.68±2.84   | 28.8±1.61    | 28.05±1.80  | 28.23±2.40   | 29.38±1.75   |

Considering the MAE and PSNR metrics, on average the eCNN method revealed better performance on the whole pelvis region followed closely by the ResNet method.

**Table 2.** Statistics of image comparison between the reference CTs and synthetic CTs generated through the different methods within air cavity, bone, and soft tissue in terms of MAE, ME, PCC, SSIM, and PSNR (average ± standard deviation).

|              | Region      | eCNN          | Atlas-based     | U-net           | GAN            | V-net            | ResNet          |
|--------------|-------------|---------------|-----------------|-----------------|----------------|------------------|-----------------|
| **MAE (HU)** | Air         | 176.35±63.11  | 607.95±95.03    | 581.09±108.44   | 198.45±50.37   | 296.87±104.88    | 195.79±104.88   |
|              | Bone        | 116.35±63.77  | 226.09±85.07    | 215.44±72.59    | 141.07±31.97   | 185.44±47.66     | 128.35±32.31    |
|              | Soft tissue | 16.41±5.42    | 68.34±20.81     | 23.51±6.32      | 27.08±4.14     | 42.66±14.08      | 21.18±5.41      |
| **ME (HU)**  | Air         | -95.65±63.06  | -375.54±189.80  | -635.51±145.05  | -83.48±77.78   | -231.00±134.05   | -115.38±85.43   |
|              | Bone        | 61.71±73.91   | -93.49±129.81   | 173.81±85.09    | 52.64±61.83    | 131.35±68.73     | 55.73±55.38     |
|              | Soft tissue | 1.53±7.60     | -53.51±28.83    | -5.58±10.37     | 4.86±14.51     | -13.32±20.12     | -0.08±12.67     |
| **PCC**      | Air         | 0.55±0.18     | 0.38±0.22       | 0.11±0.11       | 0.39±0.16      | 0.42±0.13        | 0.44±0.16       |
|              | Bone        | 0.79±0.15     | 0.63±0.16       | 0.62±0.12       | 0.75±0.03      | 0.68±0.04        | 0.76±0.04       |
|              | Soft tissue | 0.91±0.05     | 0.67±0.15       | 0.81±0.05       | 0.78±0.04      | 0.78±0.04        | 0.84±0.03       |
| **SSIM**     | Air         | 0.99±0.01     | 0.94±0.02       | 0.97±0.01       | 0.51±0.14      | 0.47±0.12        | 0.45±.08        |
|              | Bone        | 0.99±0.01     | 0.96±0.02       | 0.98±0.01       | 0.77±0.07      | 0.75±0.06        | 0.80±0.03       |
|              | Soft tissue | 0.99±0.01     | 0.96±0.02       | 0.97±0.02       | 0.90±0.08      | 0.95±0.01        | 0.95±0.04       |
| **PSNR**     | Air         | 21.91±2.88    | 11.31±2.26      | 11.58±1.88      | 19.93±1.54     | 17.41±2.63       | 19.34±2.13      |
|              | Bone        | 20.94±3.22    | 30.31±1.68      | 20.46±2.70      | 22.29±0.97     | 21.36±1.11       | 23.11±1.13      |
|              | Soft tissue | 28.78±4.90    | 23.88±3.62      | 35.98±1.46      | 29.24±2.28     | 30.85±3.44       | 31.27±2.67      |
| **DSC**      | Air         | 0.88±0.06     | 0.55±0.23       | 0.45±0.27       | 0.69±0.12      | 0.67±0.08        | 0.76±0.08       |
|              | Bone        | 0.88±0.07     | 0.76±0.06       | 0.70±0.06       | 0.83±0.03      | 0.81±0.03        | 0.84±0.03       |
|              | Soft tissue | 0.98±0.01     | 0.97±0.01       | 0.91±0.04       | 0.98±0.01      | 0.97±0.01        | 0.98±0.01       |

The axial views of the sCT and the reference CT images together with the corresponding binary masks of air cavities, bone, and soft tissue obtained from the different images are shown in Figure (8). The details of the different approaches' overall performance evaluated within the major tissue classes are tabulated in table 2. A similar trend is observed in the tissue-specific analysis of the different synthetic CT generation approaches, where, the eCNN and ResNet approaches exhibited better performance.

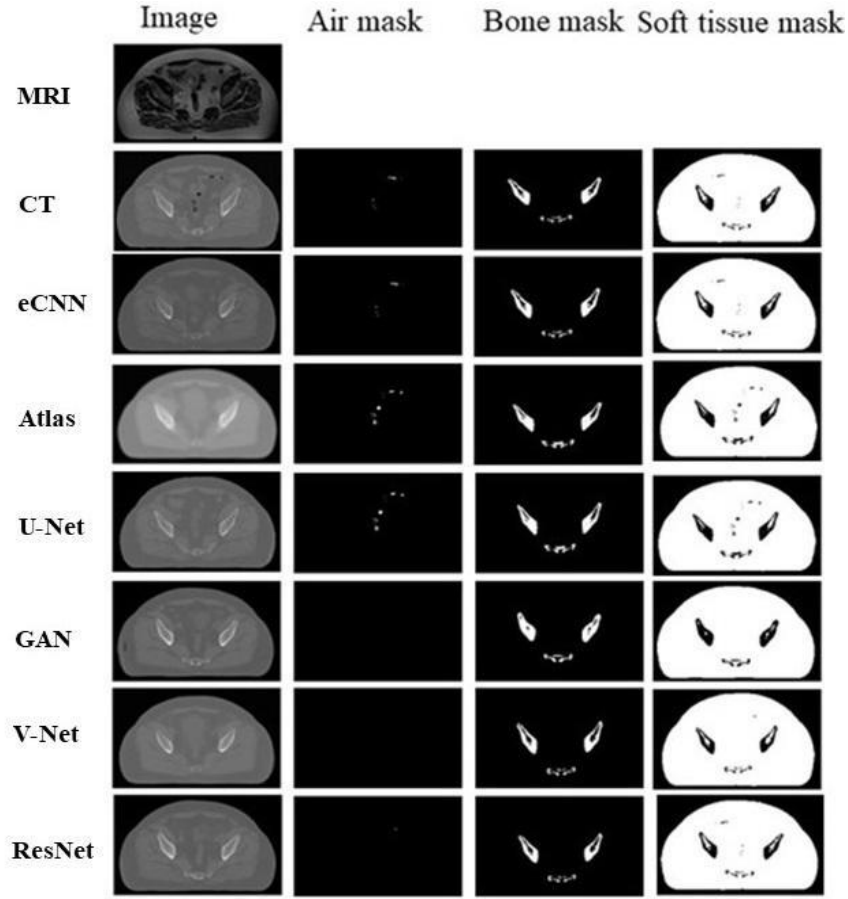

**Figure 8.** Representative slices of the sCT generated by the different methods as well as ground truth CT images together with the air cavities, bone, and soft tissue masks obtained from the different CT images.

## 4. Discussion

The promising performance of the deep learning approaches to synthesize a "pseudo-CT" from MR-only images for the tasks of MR-guided attenuation correction in PET imaging and MR-only radiation planning is evident in [22, 32, 44, 45]. Five deep learning algorithms together with an Atlas-based method were implemented/evaluated for the synthetic CT generation from pelvis MR images in this article. Comparison between the sCT image quality generated through these models and the ground truth CT indicates that both the eCNN, V-Net, and Res-Net are of the lowest noise and the highest visual inspection, while, the measured MAE, ME, PCC, and PSNR metrics indicate outperformance of eCNN and ResNet models vs. the other models. All in all, the deep learning approaches reveal acceptable performance with ranges of errors reported in the literature.

As to Table 1, the average MAE obtained through the eCNN algorithm was 26.03±8.85 HU, thus, the best result among the different methods. Because of the low sensitivity of PET attenuation correction and MR-only radiation tasks to the CT value the recorded error levels in synthetic CT estimation are tolerable in clinical practice for these tasks [3, 46]. The performance of ResNet model has close correspondence to that of the eCNN model with a MAE=31.07±7.42 HU, which, would result in intolerable errors for the PET attenuation and MR-only radiation planning tasks. Compared to available findings, the eCNN and ResNet models yield better results than those obtained by Emami et al. [47] with MAE=89.30±10.25 HU and Arabi et al. [22] with MAE=101 ± 40. The PCC for the eCNN is 0.93±.05 which surpasses all other models, indicating that the sCT generated through the eCNN model is more similar to the ground truth CTs. The ResNet model and Atlas-based approach yielded the highest and the lowest PSNR (29.38±1.75 and 23.68±2.84, respectively) indicating the effectiveness of this deep learning model. The tissue-specific performance assessment of the different approaches revealed the same trend, where, the eCNN (MAE=116.35±63.77 HU) and ResNet (MAE=128.35±32.31 HU) resulted in relatively low/tolerable errors particularly in comparison with the Atlas-based method which had the highest MAE of 226.09±85.07 HU.

In general, different deep learning architectures would lead to considerable differences in performance, thus, for any specific application different deep learning algorithms should be implemented/evaluated in determining the model with the best performance. In this article, eCNN and ResNet models revealed excellent performance in synthetic CT estimation from MR images. The ResNet model is equipped with the dilation convolutional kernels which enable it to process the input image at the highest resolution (without reducing the resolution of the input image) and extract discriminative features from the input image(s). The new building block structure, applied in the eCNN model, provided an effective update of the free parameters and a very efficient training convergence rate. For these reasons, these two models exhibited superior performance over the other deep learning models. The deep learning approaches, in general, are of better performance than the Atlas-based method.

## 5. Conclusion

Five state-of-the-art deep learning approaches together with an Atlas-based method were implemented and evaluated for the synthetic CT estimation from the MR images. The deep learning-based methods revealed high performance over the Atlas-based method. Among the deep learning approaches, the eCNN and ResNet models had the lowest quantification errors. The error levels observed in these two models would be tolerable for the accurate MR-guided PET attenuation correction and MR-only radiation planning tasks.